\documentclass[final,5p,times,twocolumn]{elsarticle} 

\usepackage{lineno,hyperref}
\usepackage{url}
\usepackage{amsfonts}
\usepackage[position=t]{subcaption}
\modulolinenumbers[5]
\journal{nim A}

\newcommand{\nech}{\ensuremath{\mathrm{Ne/(5\%CH_4)}}}
\newcommand{\arch}{\ensuremath{\mathrm{Ar/(5\%CH_4)}}}
\newcommand{\arco}{\ensuremath{\mathrm{Ar/(7\%CO_2)}}}

\biboptions{compress}

\begin{document}

\begin{frontmatter}

\title{The Resistive-Plate WELL with Argon mixtures - a robust gaseous radiation detector}

\author[WIS]{Luca Moleri}\corref{mycorrespondingauthor}\cortext[mycorrespondingauthor]{Corresponding author}\ead{luca.moleri@weizmann.ac.il}
\author[LIBPhys]{Fernando Domingues Amaro}
\author[WIS]{Lior Arazi}
\author[I3N]{Carlos Davide Rocha Azevedo}
\author[CERN]{Eraldo Oliveri}
\author[WIS]{Michael Pitt}
\author[WIS]{Jana Schaarschmidt}
\author[WIS]{Dan Shaked-Renous}
\author[LIBPhys]{Joaquim Marques Ferreira dos Santos}
\author[I3N]{Jo\~{a}o Filipe Calapez de Albuquerque Veloso}
\author[WIS]{Amos Breskin}
\author[WIS]{Shikma Bressler}

\address[WIS]{Department of Particle Physics and Astrophysics, Weizmann Institute of Science, 76100 Rehovot, Israel}
\address[LIBPhys]{LIBPhys, Department of Physics, University of Coimbra, RuaLarga, PT3004-516 Coimbra, Portugal}
\address[I3N]{I3N, Physics Department, University of Aveiro, 3810-193 Aveiro, Portugal}
\address[CERN]{CERN, Meyrin, Switzerland}

\begin{abstract}
A thin single-element THGEM-based, Resistive-Plate WELL (RPWELL) detector was operated with 150~GeV/c muon and pion beams in \nech, \arch~ and \arco; signals were recorded with 1~cm$^2$ square pads and SRS/APV25 electronics.  Detection efficiency values greater than 98\%  were reached in all the gas mixtures, at average  pad multiplicity of 1.2. The use of the 10$^9\Omega$cm resistive plate resulted in a completely discharge-free operation also in intense pion beams. The efficiency remained essentially constant at 98-99$\%$ up to fluxes of $\sim$10$^4$Hz/cm$^2$, dropping by a few $\%$ when approaching 10$^5$~Hz/cm$^2$. These  results  pave  the way towards cost-effective, robust, efficient, large-scale  detectors  for  a variety of applications in future particle, astro-particle and applied fields. A potential target application is digital hadron calorimetry.
\end{abstract}


\end{frontmatter}


\section{Introduction}
The Thick Gas Electron Multiplier (THGEM) is a robust radiation detection element suitable  for applications requiring large detection areas~\cite{chechik2004thick,breskin2009concise}. The  broad  interest in THGEM-based detectors has resulted in the development of production techniques and concepts (e.g. \cite{liu2013successful,alexeev2013ion}), including the use of resistive films and materials for reducing occasional discharge effects~\cite{peskov2007development, charpak2009progress, bressler2013recent}. In  this  context, the  experience  gained  with  various  configurations  of  THGEM-based  multipliers  with  resistive  anodes~\cite{bressler2013recent,arazi2012thgem,arazi2014laboratory} has led to  the development of a particularly promising candidate - the Resistive-Plate WELL (RPWELL)~\cite{rubin2013first}. It is a single-sided copper-clad THGEM electrode, coupled to a segmented readout anode (pads or strips) through a thin high bulk-resistivity ($\sim$10$^8$-10$^{10}$~$\Omega$cm) plate.
Extensive laboratory studies in \nech~ demonstrated discharge-free operation at high gas-avalanche gains and over a broad ionization range~\cite{rubin2013first}. Good performances in terms of efficiency and average pad multiplicity - studied in the context of future (Semi) Digital Hadronic Calorimeter ((S)DHCAL)~\cite{behnke2013international}, were reached, in \nech~ with 150~GeV/c muon and pion beams using a thin 10$\times$10 cm$^2$ RPWELL, with a resistive anode (Semitron$^{\circledR}$ ESD225) of bulk resistivity of $\sim$10$^9$$\Omega$cm~\cite{bressler2016first}.
In the present article we report on new results of further beam studies of this detector, extended to the low-cost Ar-based gas mixtures \arch~ and \arco. Detection efficiency values of $\sim$98\% were reached in all conditions, at average pad multiplicity of $\sim$1.2, in discharge-free operation, also with a high-intensity pion flux. The detector fulfill the requirements for sampling elements in (S)DHCAL and has the potential for applications requiring robust, cheap, efficient large-scale detectors with moderate spacial and energy resolutions.

\section{Experimental  setup  and  methodology}
\subsection{RPWELL detector, tracking and readout system}
\begin{figure}[h]
\begin{subfigure}{0.6\linewidth}\caption{}
\includegraphics[scale=0.2]{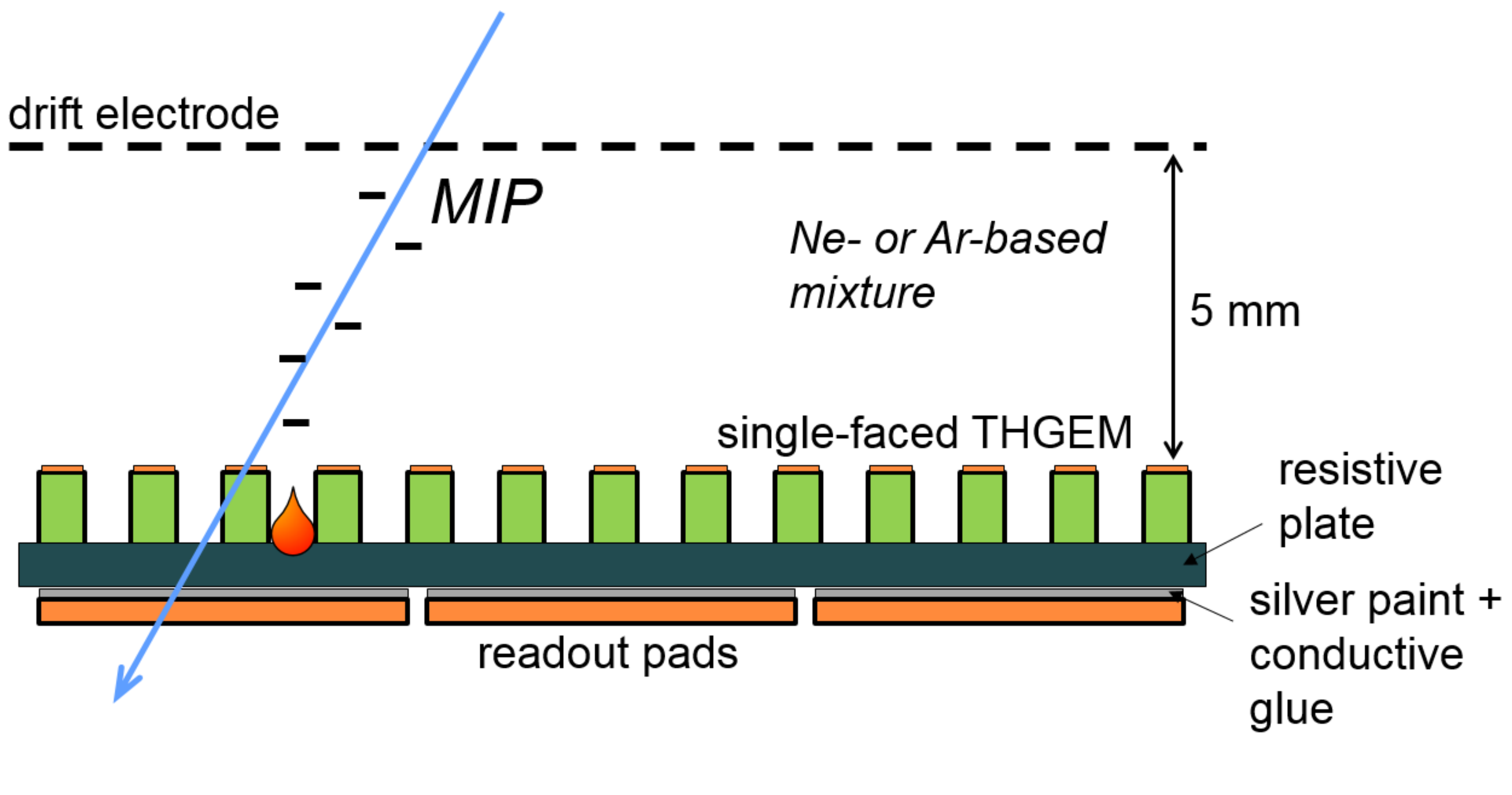}
\end{subfigure}\vfill
\begin{subfigure}{0.3\linewidth}\caption{}
\includegraphics[scale=0.6]{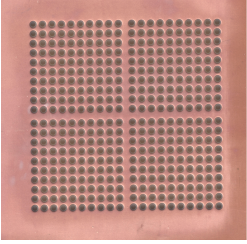}
\end{subfigure}
\begin{subfigure}{0.2\linewidth}\caption{}
\includegraphics[scale=0.075]{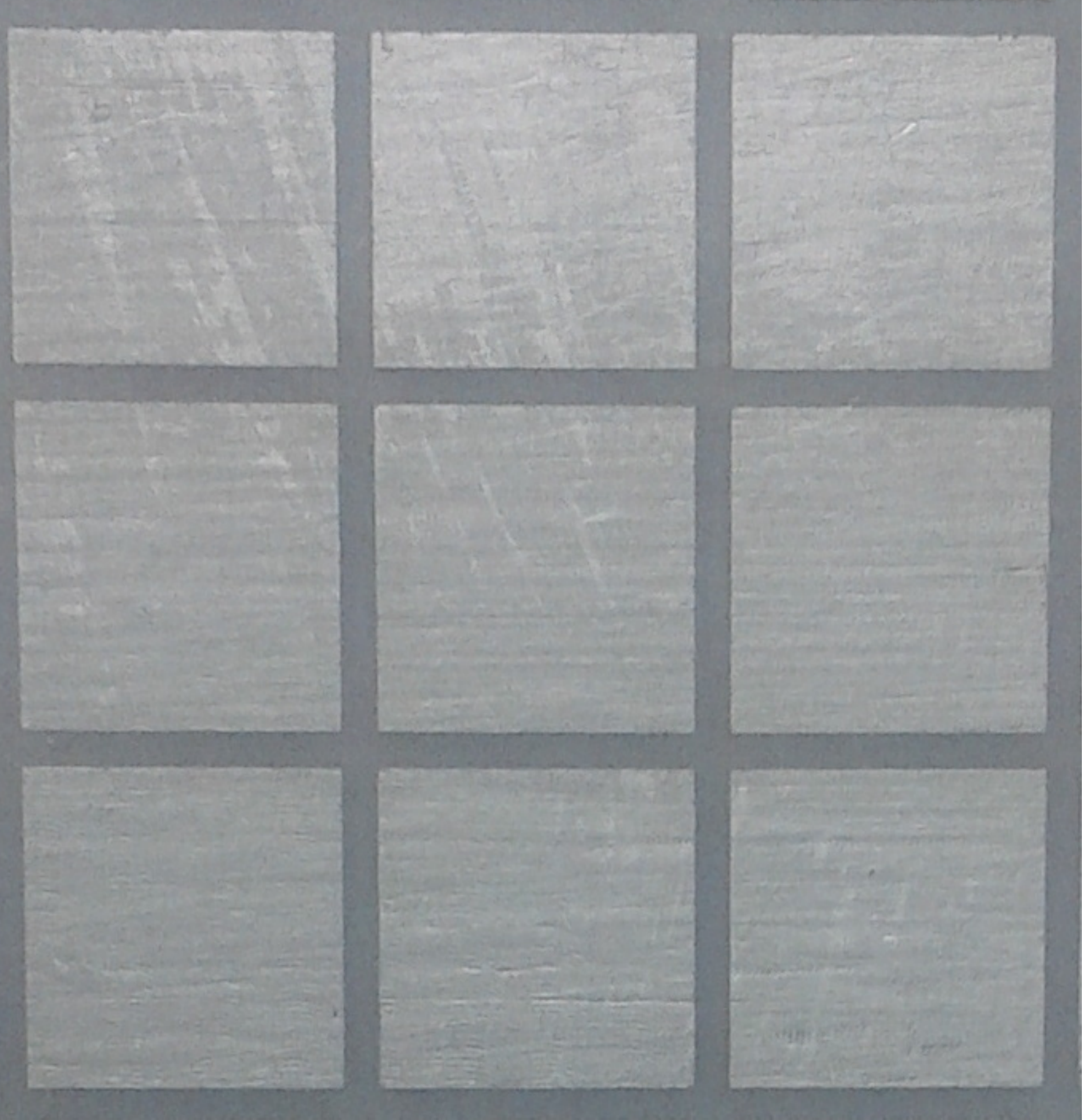}
\end{subfigure}
\begin{subfigure}{0.2\linewidth}\caption{}
\includegraphics[scale=0.1]{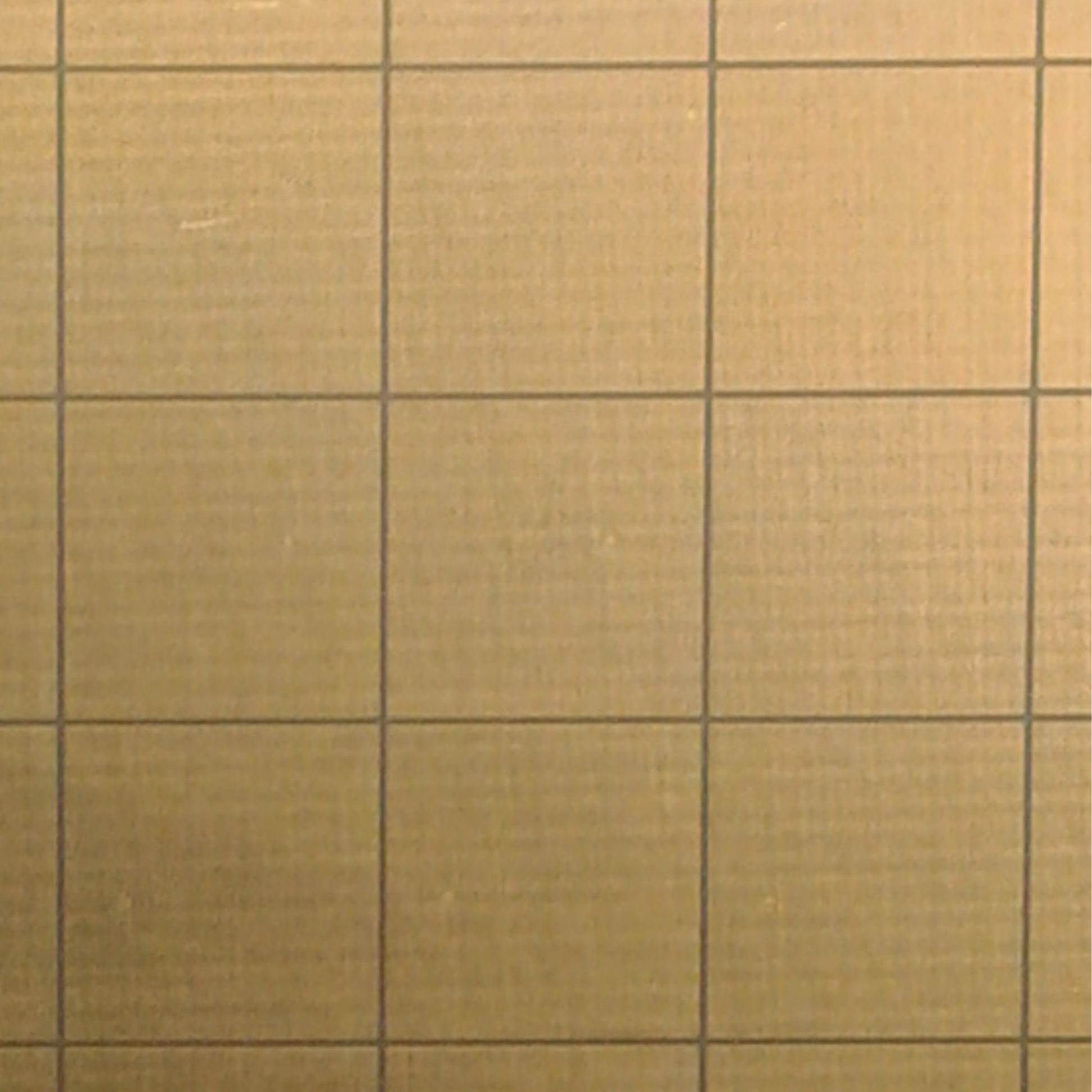}
\end{subfigure}
\caption{The 10$\times$10~cm$^2$ RPWELL detector scheme. The RPWELL structure (a): a single sided THGEM (b) coupled to the readout anode through a resistive plate. The anode readout pads (d) are coupled to conductive pads patterned on the resistive plate (c). The metal bands in (b) are located above the underlying pad borders (c,d).}\label{fig: detector scheme}
\end{figure}

The 10$\times$10~cm$^2$ RPWELL detector, its SRS/APV25 readout system~\cite{martoiu2013development,french2001design} and the experimental setup at the CERN-SPS/H4 beam-line, were detailed in~\cite{bressler2016first} and are briefly described here. The detector scheme, elements and operation principle are shown in Fig.~\ref{fig: detector scheme}. The single-sided THGEM  electrode, 0.86~mm thick, had 0.5~mm diameter holes mechanically drilled in an FR4 plate, copper-clad on one side. The holes were arranged in a square lattice (Fig.~\ref{fig: detector scheme}-b), with 0.96~mm pitch, so that they cover the underlying 1$\times$1~cm$^2$ anode pads, but not their borders, where 0.86~mm wide metal bands are left, as described in~\cite{arazi2012thgem}. The plate was chemically etched, yielding 0.1~mm rims around the holes, preventing sharp edges and eventual defects. The  THGEM electrode was coupled to the anode pads (Fig.~\ref{fig: detector scheme}-d) through a 0.4~mm thick Semitron$^{\circledR}$ ESD225\footnote{www.quadrantplastics.com} static dissipative plastic plate  (2$\times$10$^9$~$\Omega$cm bulk resistivity). The electrical contact between the resistive plate and the readout pads is essential for efficient clearance of the avalanche electrons. Therefore, the bottom of the resistive material was patterned  with conductive pads (Fig.~\ref{fig: detector scheme}-c), individually connected to the anode pads (like in~\cite{bressler2016first} but using 3M\textsuperscript{TM} Electrically Conductive Adhesive Transfer Tape 97073\footnote{www.3m.com}).

The detector electrodes  were individually biased using CAEN A1833P and A1821N HV power-supply boards, remotely  controlled  with  a CAEN SY2527 unit. The voltage  and current in each channel were  monitored and  stored. All HV  inputs were connected through low-pass filters. The RPWELL bias ($\Delta$V$_{RPWELL}$)  with  respect to the grounded anode was  varied throughout the experiment, while the drift voltage was kept constant: $\Delta$V$_{drift}$= 250~V - corresponding to a  drift field of $\sim$0.5~kV/cm across  the 5~mm drift gap. The detector was operated in \arch~ and \arco~ gas mixtures at atmospheric pressure and room temperature, in gas-flow mode (50-100 cc/min); measurements in  \nech~ were taken for comparison.
The operation in the Ar mixtures required higher voltages compared to the ones in \nech~ to obtain similar gains. However, with respect to Ne, Ar has the advantage of larger number of electron-ion pairs produced by a minimum ionizing particle (at normal conditions 94 compared to 39 e-i pairs/cm~\cite{sauli2014gaseous}) - allowing for a smaller drift gap for equal number of primary charges - and considerably lower cost (which might be crucial for large-scale systems).
The use of CO$_2$ instead of CH$_4$ as a photon quencher is preferable because the former is non flammable.
The triggering, tracking and DAQ system was the same as described in~\cite{bressler2016first}. The RPWELL chambers were placed along the beam line in-between the tracker elements.

\subsection{Working point: $\Delta$V$_{RPWELL}$, threshold and matching parameter} 
As described in~\cite{bressler2016first}, for each event the detector pads with signal above threshold were grouped into clusters of neighbors. The global detection efficiency was then calculated as the fraction of particle tracks matched to a cluster in the detector, while the average pad multiplicity was the average  number of pads contained in each cluster. The detector working point was adjusted to optimize its performance in each gas mixture, targeting high global detection efficiency at low average pad multiplicity. The lowest-possible value (closest to 1) of the latter is a prerequisite for valid particle counting, e.g. in a potential application as a sampling element in DHCAL~\cite{arazi2012thgem,sefkow2015experimental}. 
The optimization was done using a set of measurements with $\sim$100~Hz/cm$^2$ wide (5$\times$5~cm$^2$) muon beam and a $\sim$13000~Hz/cm$^2$ narrow (2$\times$2~cm$^2$) pion beam. In both cases, only tracks hitting the detector in a 4$\times$4~cm$^2$ central region were considered. Two important parameters in the analysis are the  the threshold for zero order suppression (ZSF) and the track-cluster matching parameter (W) (see details in~\cite{bressler2016first}). The values of ZSF and W were fixed at ZSF= 15 and W= 10 or 15~mm following the same method described in~\cite{bressler2016first}.

\section{Results}
\subsection{Detected charge, global detection efficiency and average pad multiplicity}

\begin{figure}[!h]
\includegraphics[scale=0.32]{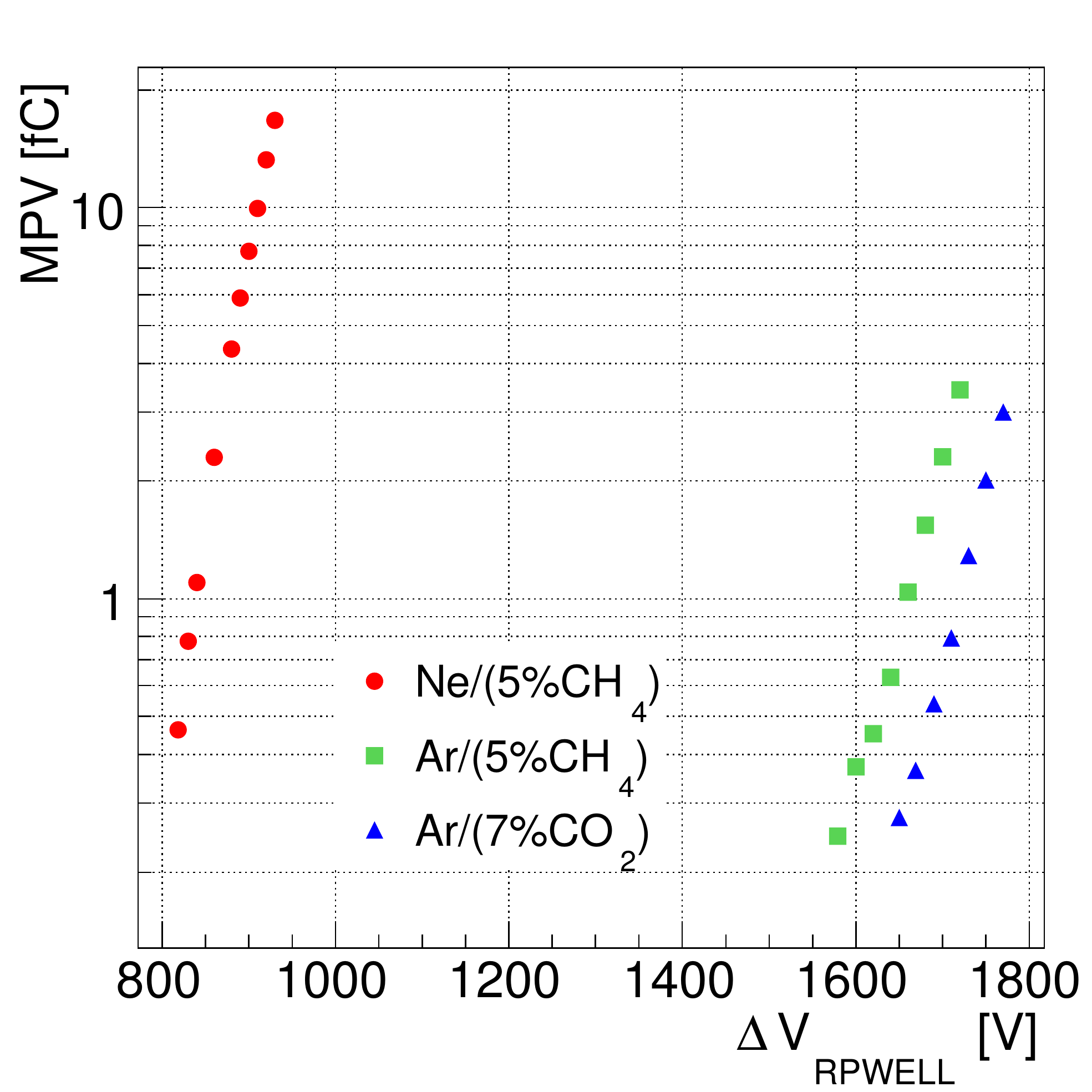}
\caption{The most probable value (MPV) of the charge measured by the RPWELL detector in $\sim$100~Hz/cm$^2$ muon beam for different $\Delta$V$_{RPWELL}$ values in the three gas mixtures.}\label{fig: HV scan - gain}
\end{figure}

\begin{figure}[!h]
\includegraphics[scale=0.32]{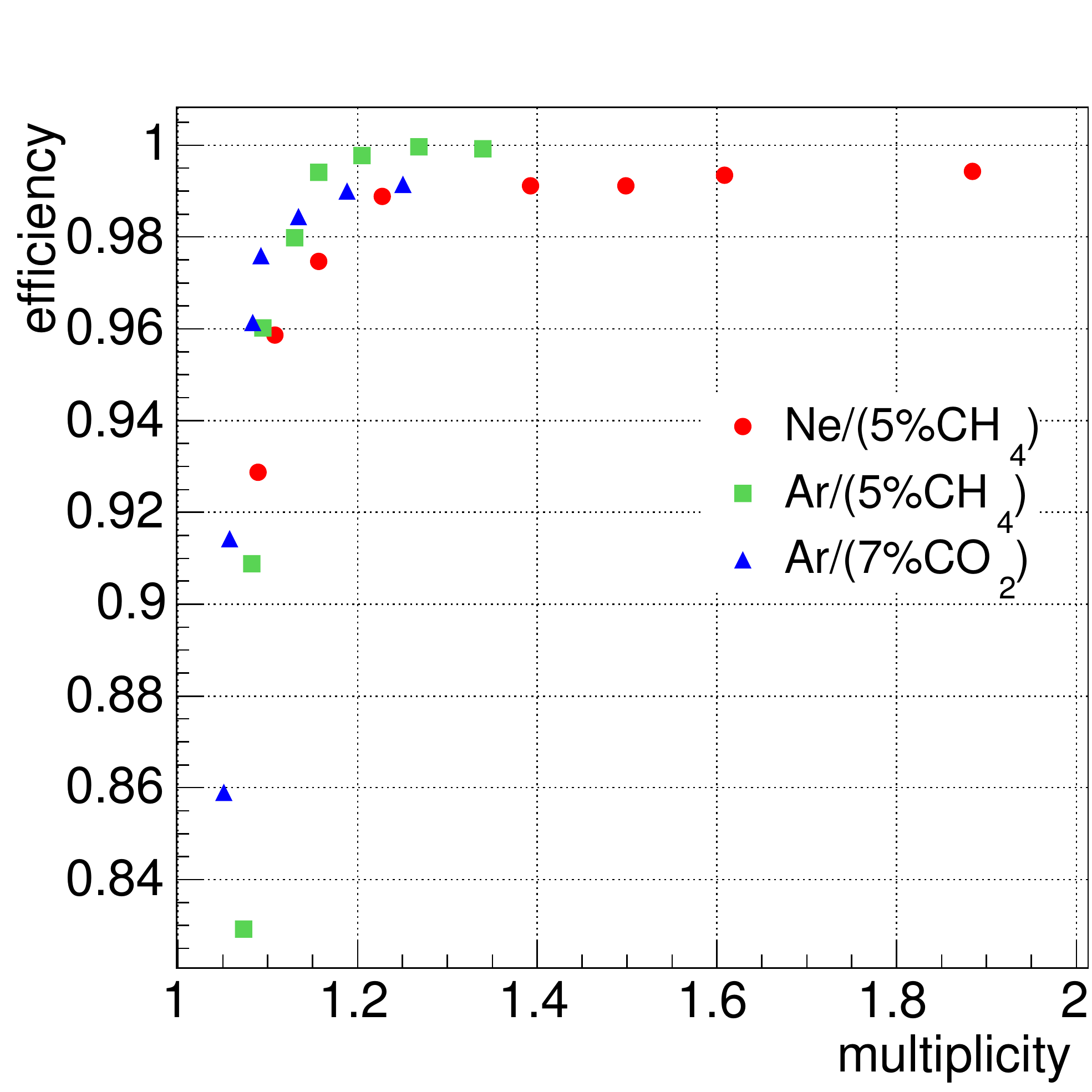}
\caption{The global detector efficiency versus average pad multiplicity of the RPWELL detector in $\sim$100~Hz/cm$^2$ muon beam in the three gas mixtures (using the same data set shown in Fig.~\ref{fig: HV scan - gain}).}\label{fig: HV scan - eff}
\end{figure}

We studied the detector operation in \nech, \arch~ and \arco~ using $\sim$100~Hz/cm$^2$ muon beam.
Fig.~\ref{fig: HV scan - gain} depicts the most probable value (MPV) of the detected charge, derived from the recorded Landau distributions as a function of $\Delta$V$_{RPWELL}$. It is important to note that the effective measured charge is a few times smaller than the avalanche charge; this results from the convolution of the long signal rise time in the RPWELL ($\sim$1-2~$\mu$s rise-time) and the short shaping time of the APV25 chip ($\sim$75~ns) - as described in~\cite{bressler2016first}. Fig.~\ref{fig: HV scan - eff} shows the global detector efficiency as a function of the average pad multiplicity in the same data set shown in Fig.~\ref{fig: HV scan - gain}. $\Delta$V$_{RPWELL}$ values of 880~V, 1640~V and 1750~V in \nech, \arch, and \arco~ respectively, gave very high detection efficiency (98-99$\%$) and low pad multiplicity ($\sim$1.2) in a stable condition.

\subsection{Performance under low and high particle flux}
\label{sec: rate scan}
\begin{figure}
\includegraphics[scale=0.32]{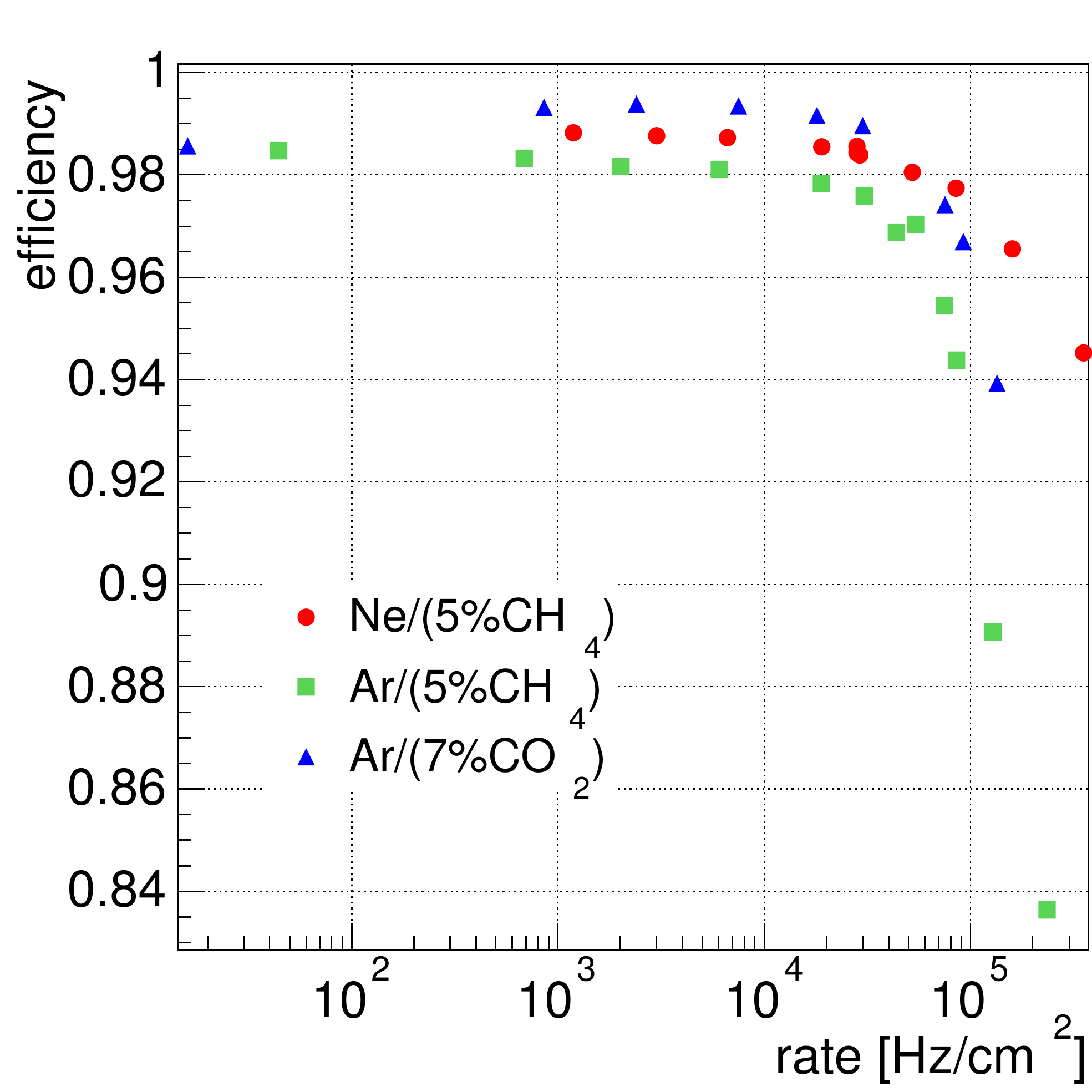}
\caption{Global detection efficiency of the RPWELL detector as a function of the incoming particle flux in \nech, \arch~ and \arco. The values of $\Delta$V$_{RPWELL}$ were 880~V, 1700~V and 1770~V respectively.}
\label{fig: Rate scan}
\end{figure}

The RPWELL performance was investigated with low-rate muon and high rate pion beams reaching a flux of $\sim$4$\cdot$10$^5$~Hz/cm$^2$. The results are shown in Fig.~\ref{fig: Rate scan}. Note that in order to keep a high efficiency at high particle fluxes, these measurements were done using higher voltages than those optimal for detecting low-rate muons. The values of $\Delta$V$_{RPWELL}$ were 880~V, 1700~V and 1770~V in \nech, \arch~ and \arco~ respectively. The global detection efficiency (Fig.~\ref{fig: Rate scan}) is stable until rates of $\sim$10$^4$Hz/cm$^2$, consistently for all three gas mixtures. It drops by a few \% (to 94\%) while approaching rates of $\sim$10$^5$Hz/cm$^2$, due to 30\% gain loss (not shown), possibly resulting from the charging up of the holes and avalanche build-up limitations on the resistive anode (see for example~\cite{affatigato2015measurements}). These results are in agreement with that previously shown in \nech~\cite{bressler2016first}. The efficiency drop can be mitigated using higher operation voltage.

\begin{figure}[!h]
\includegraphics[scale=0.35]{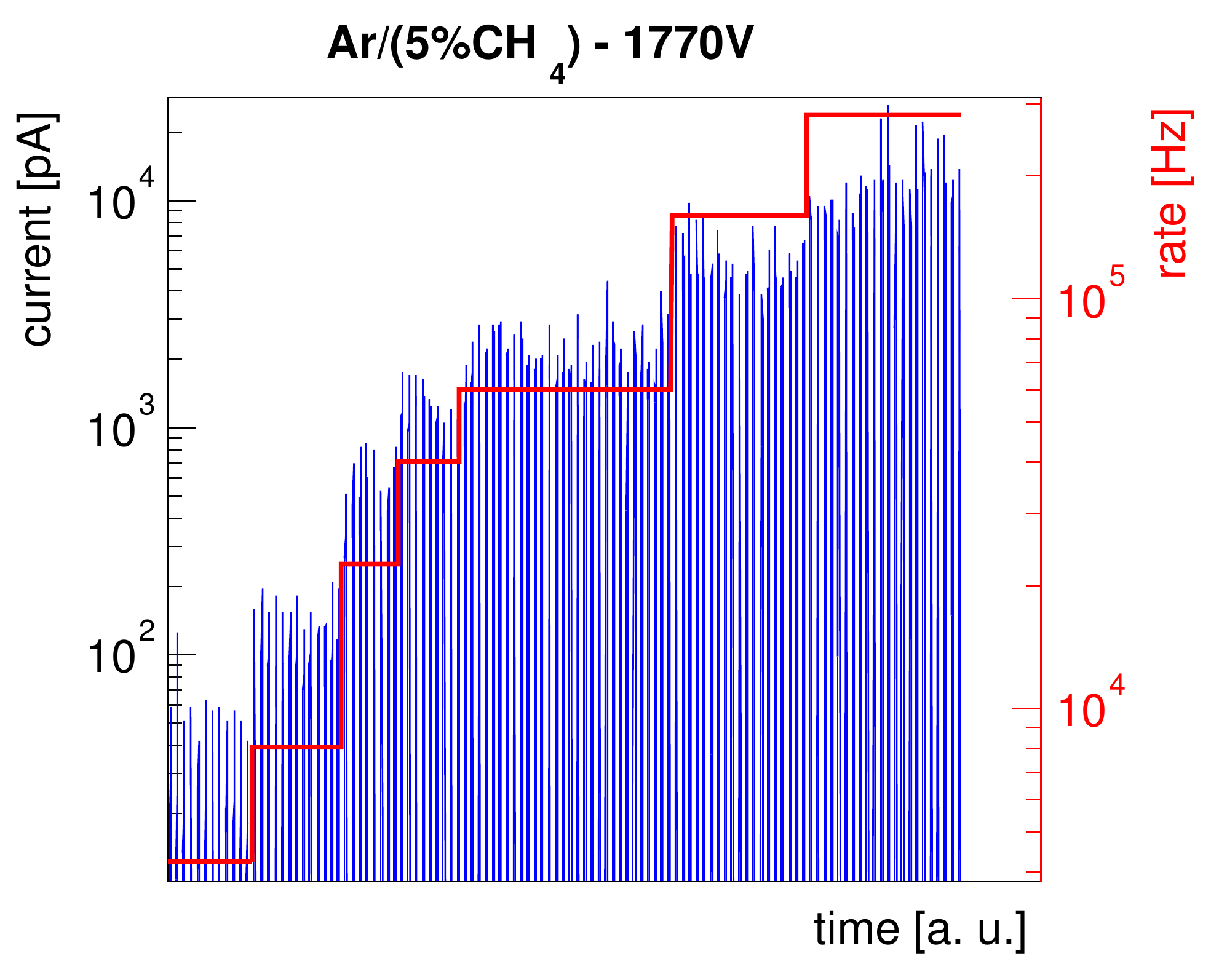}
\caption{Current flowing through the detector during pion runs at different rates in \arch. The beam spill-structure is clearly visible.}
\label{fig: femtometer - readings}
\end{figure}

To demonstrate the electrical stability of the RPWELL, we measured the current flowing through the anode using a sensitive ammeter~\cite{Femtobox}, while irradiating the detector with pions at different rates. Fig.~\ref{fig: femtometer - readings} shows the current and the pion rates as a function of time. The measurement shown is in \arco, and similar results were obtained in all three gas mixtures. As expected, the small current spikes, corresponding to the beam spill-structure, grow smoothly in amplitude with the particle rate. The measured value of the current should follow the simple expression: $\mathrm{I= q\cdot n\cdot \Phi\cdot G(\Phi)}$, where I is the current, q is the electron charge, n is the number of electron-ion pairs produced by a minimum ionizing particle in 5~mm of Ar~\cite{sauli2014gaseous}, $\mathrm{\Phi}$ is the particle rate and G the detector gain (which depends on $\Phi$ as explained above). Reversing the formula and using the measured value of the current, we can estimate the value of G. For example for $\mathrm{\Phi}$=4$\cdot$10$^3$~Hz we get G= 1.3$\times$10$^3$, which is compatible with the value obtained from the measured charge MPV (Fig.~\ref{fig: HV scan - gain} which is taken at a similar particle rate), once taking into account the effect of the electronics shaping time.

\subsection{Gain stability over time}
\label{sec: stability}
\begin{figure}[!h]
\includegraphics[scale=0.32]{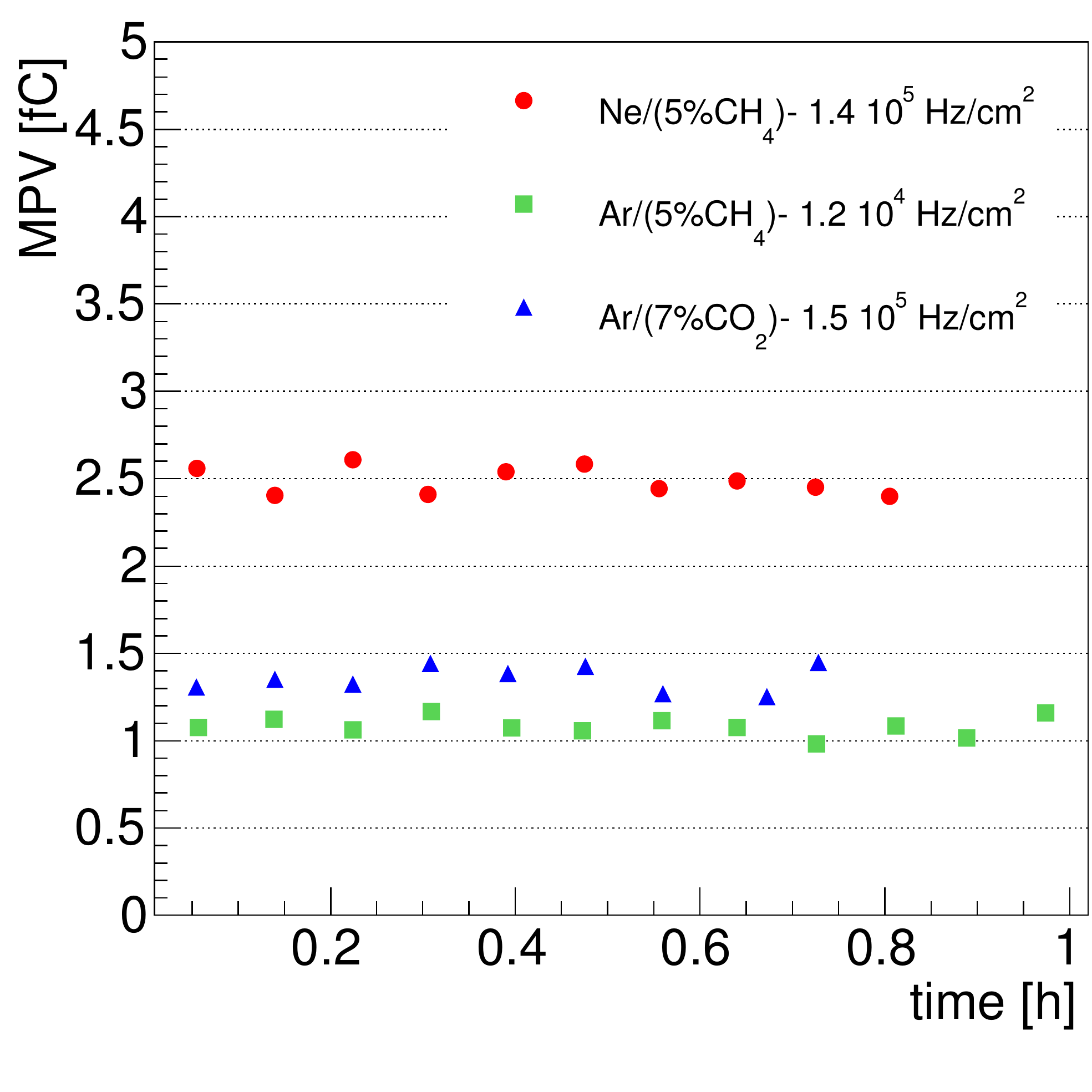}
\caption{Gain stability over time under a high-rate (10$^4$-10$^5$~Hz/cm$^2$) pion flux in \nech, \arch~ and \arco. The values of $\Delta$V$_{RPWELL}$ were 880~V, 1700~V and 1770~V respectively.}
\label{fig: Gain stability}
\end{figure}

A stable operation of the detector was demonstrated over time, under 10$^4$-10$^5$~Hz/cm$^2$ pion fluxes, as shown in Fig.~\ref{fig: Gain stability}. The applied voltages were the same as those of the measurements presented in section~\ref{sec: rate scan}. No significant gain variations (less than $5\%$) were observed along  $\sim$1 hour of operation in all three gas mixtures. The values of global detection efficiency and average pad multiplicity during these measurements also remained stable.
 
\subsection{Discharge probability}

A discharge was defined as an abrupt increase in the current supplied to the detector~\cite{bressler2016first}. Discharge probability was measured during the high-rate pion runs presented in section~\ref{sec: stability}. 
No discharges were observed in any of the gas mixtures while irradiating the detector with over 10$^8$ pions; therefore the resulting value of 10$^{-8}$ is an upper limit for the discharge probability in the present RPWELL configuration. Since pions are prone to induce highly-ionizing secondary events, we have an additional indication on the broad dynamic range of this detector. 

\section{Summary and discussion}
 
A 6~mm thick (without readout) 10$\times$10 RPWELL detector with a Semitron$^{\circledR}$ ESD225 resistive plate coupled to a pad readout anode was investigated for the first time in \arch~ and \arco~ gas mixtures; its performance was compared with that in \nech. This thin, single-stage detector was operated with 150~GeV/c muons and pions, at fluxes reaching 4$\cdot$10$^5$~Hz/cm$^2$. High detection efficiency values, greater than 98$\%$, at low average pad multiplicity of $\sim$1.2, were demonstrated in all three gas mixtures, maintaining stable, discharge-free operation, also at high pion flux. The efficiency remained unaffected up to a pion flux of 10$^4$~Hz/cm$^2$, above which it decreased by a few \% at $\sim$10$^5$~Hz/cm$^2$. For example, in \arco~ the RPWELL maintained a global detection efficiency of 95$\%$ under a flux of 1.5$\cdot$10$^5$~Hz/cm$^2$ pions. The lack of electrical instabilities over more than 10$^8$ pion events sets an upper limit of 10$^{-8}$ on the discharge probability - $\sim$ 2 orders of magnitude better than other THGEM-based configurations (e.g. single- and double-THGEM, Resistive WELL and Segmented Resistive WELL reviewed in~\cite{bressler2013recent}).

Compared to other detector technologies explored under similar conditions, e.g. for the DHCAL, the performance of the 10$\times$10~cm$^2$ RPWELL detector, with respect to detection efficiency and pad multiplicity, is superior to that of 1$\times$1~m$^2$ RPCs~\cite{affatigato2015measurements,bilki2008calibration} and 30$\times$30~cm$^2$ GEM detectors~\cite{yu2012application}; it is similar to that of 1$\times$1~m$^2$ MICROMEGAS~\cite{adloff2009micromegas,chefdeville2014micromegas}.
The efficiency dependence on the incoming particle flux is similar to that of multi-gap RPCs based on semi-conductive glass~\cite{naumann2011ceramics}. Also in terms of electrical stability the performance of the RPWELL detector is similar to that of RPCs.
These results pave the way towards robust, efficient large-scale detectors for applications requiring economic solutions at moderate spatial and energy resolutions. A different application than “digital counting” (like in DHCAL or RICH) could be that of a tracking detector with strip readout. In this case it would require charge spreading over a resistive layer to increase the spatial resolution (see for example~\cite{cortesi2007investigations}). This detector configuration, as well as the RPWELL technology scale-up and the investigation of other suitable resistive materials are the subject of current R\&D.

\section*{Acknowledgments}
Research supported by: I-CORE Program of the Planning and Budgeting Committee and The Israel Science Foundation (grant NO 1937/12); Nella and Leon Benoziyo Center for High Energy Physics; Horizon 2020 research and innovation program grant 654168. A. Breskin is the W.P. Reuther Professor of Research in the Peaceful use of Atomic Energy. F. D. Amaro and C. D. R. Azevedo acknowledge support by FCT under Post-Doctoral Grant SFRH/BPD/74775/2010 and SFRH/BPD//79163/2011. Work done within the CERN RD51 framework.


\bibliography{bibliography_etal.bib}

\end{document}